\documentclass[review]{elsarticle}
\pdfoutput=1
\usepackage{lineno,hyperref}
\modulolinenumbers[5]

\journal{Journal of \LaTeX\ Templates}

\begin{document}

\begin{frontmatter}

\title{RF simulation platform of qubit control using FDSOI technology for quantum computing}

\author[mymainaddress]{H. Jacquinot\fnref{myfootnote}}
\author[mysecondaryaddress]{R. Maurand}
\author[mysecondaryaddress]{G. Troncoso Fernandez Bada}
\author[mymainaddress]{B. Bertrand}
\author[mymainaddress]{M. Cassé}
\author[mysecondaryaddress]{Y. M. Niquet}
\author[mysecondaryaddress]{S. de Franceschi}
\author[mytertiaryaddress]{T. Meunier}
\author[mymainaddress]{M. Vinet}

\cortext[mycorrespondingauthor]{Corresponding author}
\ead{helene.jacquinot@cea.fr}
\address[mymainaddress]{CEA-Leti, Univ. Grenoble Alpes, F-38000 Grenoble, France}
\address[mysecondaryaddress]{Univ. Grenoble Alpes, CEA, IRIG/DEPHY, Grenoble, France}
\address[mytertiaryaddress]{CNRS Institut Néel, Grenoble, France}

\begin{abstract}
In this paper, we report on simulations of an Electron Spin Resonance (ESR) RF control line for semiconductor electron spin qubits. The simulation includes both the ESR line characteristics (geometry and configuration, stack and material properties) and the electromagnetic (EM) environment at the vicinity of the qubits such as gates and interconnect network. With the accurate assessment of the magnetic and electric field distribution, we found that the EM environment of the qubits contributes significantly to the ESR line efficiency for spin control characterized by the magnetic over electric field ratio generated at the qubit location.
\end{abstract}


\end{frontmatter}

\section{Introduction}
Thanks to their long coherence time and their compatibility with advanced semiconductor manufacturing, electron spin qubits are expected to bring breakthrough in Quantum computing technologies \cite{maurand_cmos_2016,vinet_material_2021,petit_universal_2020}.
To enable fabrication of a multi-qubits demonstrator, spin control modules need to be developed together with the qubits full integration flows. 
Spin qubit control can be achieved by electron spin resonance (ESR) \cite{poole_electron_1967}. It consists in applying to the qubit a resonant AC magnetic field generated by the AC current flowing through an RF line at the vicinity of the qubit \cite{koppens_driven_2006,dehollain_nanoscale_2013,jones_logical_2018}.

Usually, the ESR RF line is simulated without considerating the electromagnetic (EM) effects of the surroundings of the qubit such as interconnects, dummies and gate structure.
Here the simulations aim at describing a realistic environment including the qubits and a real BEOL (Back-end-of-Line) process in a FDSOI technology operating at cryogenic temperature \cite{niquet_challenges_2020,hutin_mos_2021}.

\section{ESR line FoM}
For quantum computing, the main figure-of-merits (FoM) of the ESR line we consider are:
\begin{itemize}
\item Magnetic field, ratio B/E
\item Impedance matching, dissipated power, conversion efficiency
\item Scalability, fields homogeneity
\end{itemize}

\begin{figure}[!h]
\centering
\includegraphics[width=0.7\columnwidth]{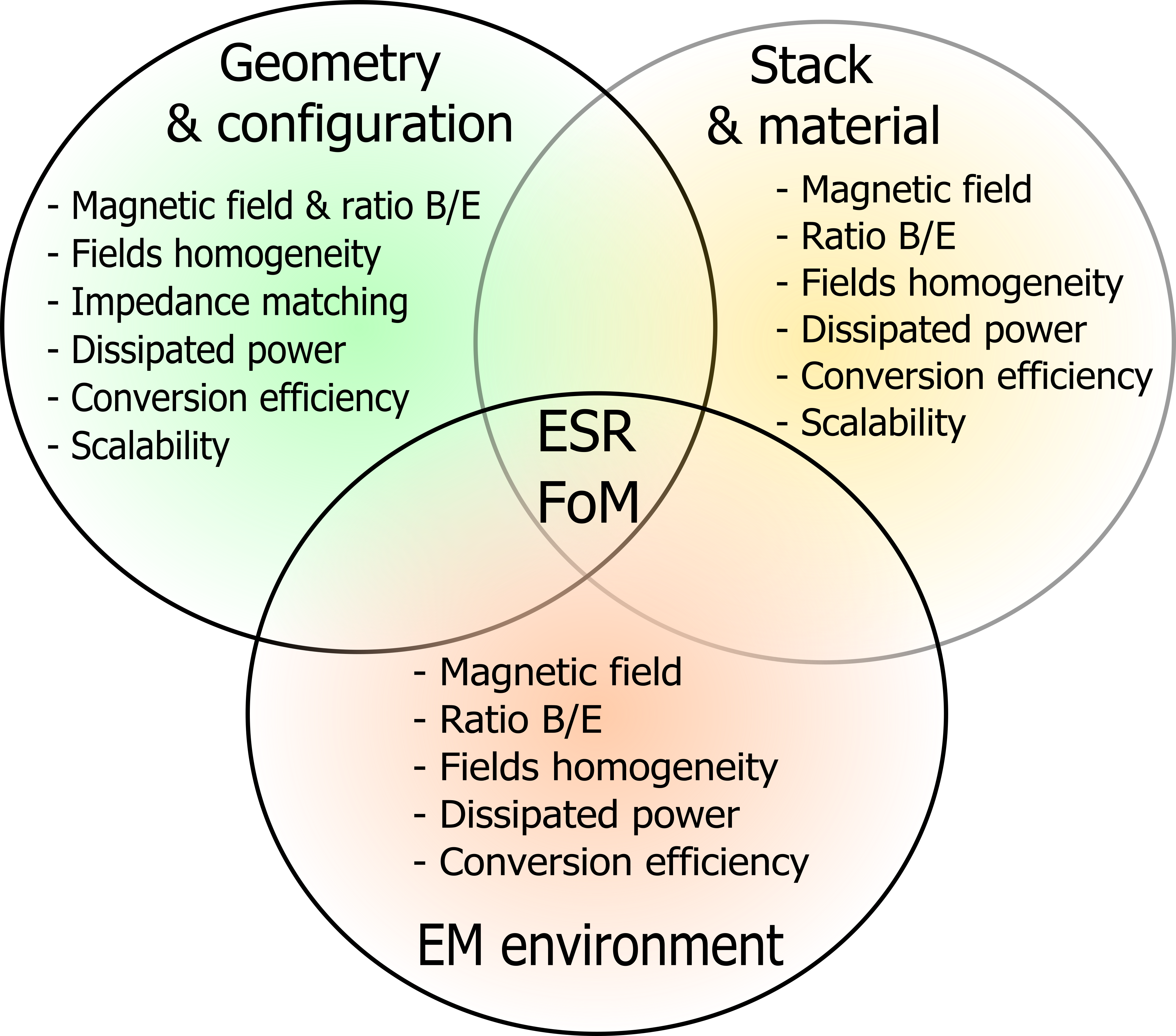}
\caption{Classification of ESR FoM: ESR line geometry and configuration, stack and material and EM environment dependent.}
\label{fig_ESR_FOM}
\end{figure}

In this paper, we propose a classification of the ESR line FoM (Fig.\,\ref{fig_ESR_FOM}), and we demonstrate that the $B$/$E$ ratio is dependent on ESR line geometry and configuration, stack and material and finally EM environment.

We will focus on the magnetic field ($B$) and the magnetic over electric field ratio ($B$/$E$) for a given ESR line input power, since the AC magnetic field is the one used for ESR (directly proportional to the spin Rabi frequency), whereas $E$ field is the parasitic one, potentially heating the sample and leading to qubits improper operations \cite{koppens_driven_2006,dehollain_nanoscale_2013,zwerver_qubits_2021}. 

\section{Electromagnetic simulation platform}

\subsection{Methodology}
In this work, we develop an electromagnetic simulation platform to assess accurately ESR line FoMs.
Simulations are realized using HFSS from Ansys and CST from Dassault System, which are finite element EM solvers, for coping with both ESR and quantum dots (QDs) co-design and multi-scale requirements.
ESR line/QDs co-simulation allows describing a realistic EM environment at the vicinity of the QDs, accounting of both interconnect lines and specificities of materials and processes, while multi-scale electromagnetic (EM) simulation aims to cover the nanometer single qubit up to the millimeter access lines interconnect.
We can then use this EM simulation platform to study the $B$/$E$ ratio FoM according to the classification proposed in Fig.\,\ref{fig_ESR_FOM}.

\begin{figure}[!h]
\centering
\includegraphics[width=0.7\columnwidth]{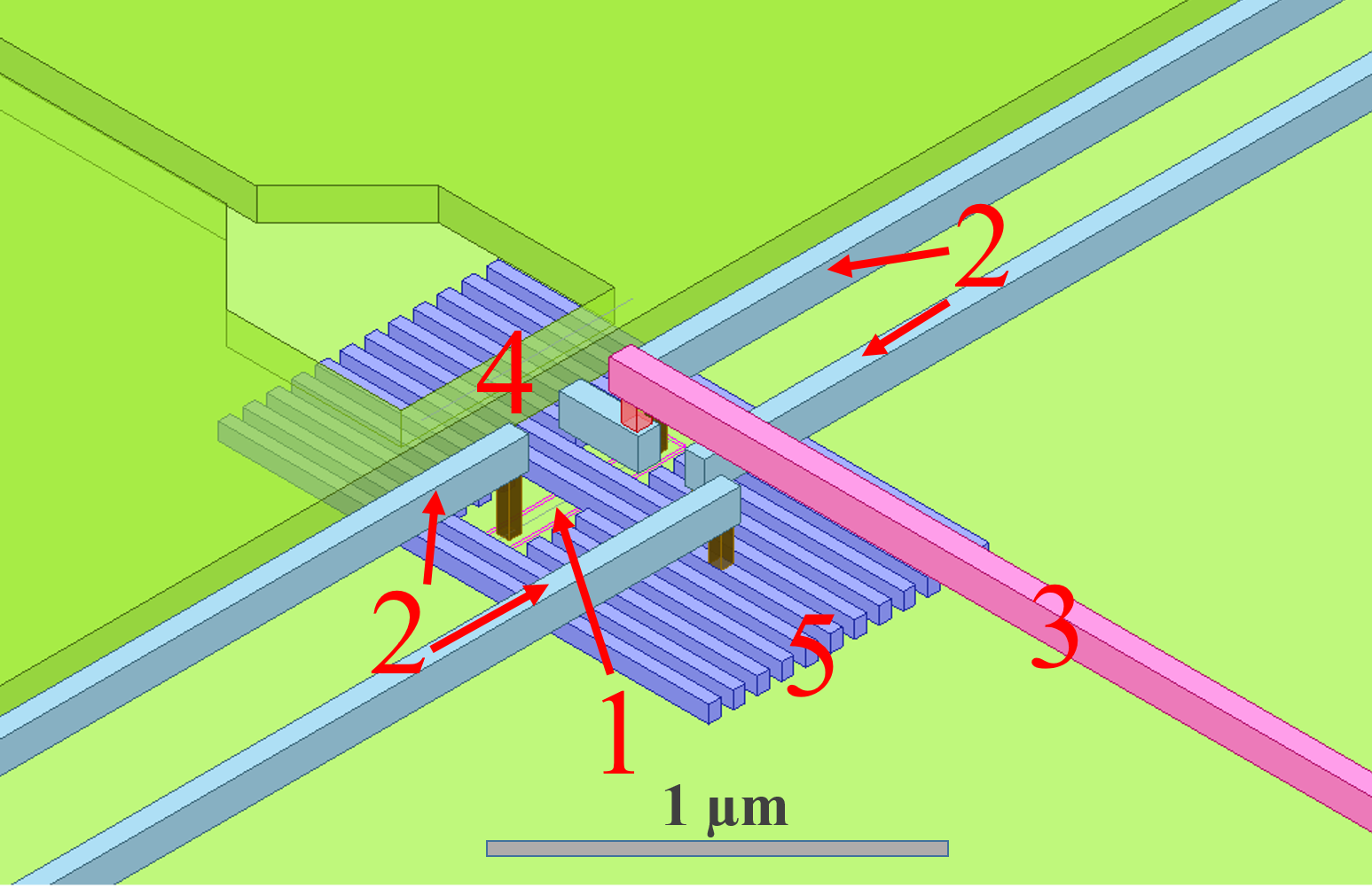}
\caption{Structure used for 3D electromagnetic (EM) simulations to assess RF performances for spin manipulation using Electron Spin Resonance (ESR) 1) silicon active, QD, 2) Gates and reservoirs, 3) Top gate, 4) ESR line nano-antenna in M1 level, 5) Polysilicon gates and dummies, illustration from \cite{niquet_challenges_2020}.}
\label{fig_ESR_dum}
\end{figure}

\begin{figure}[!h]
\centering
\includegraphics[width=0.75\columnwidth]{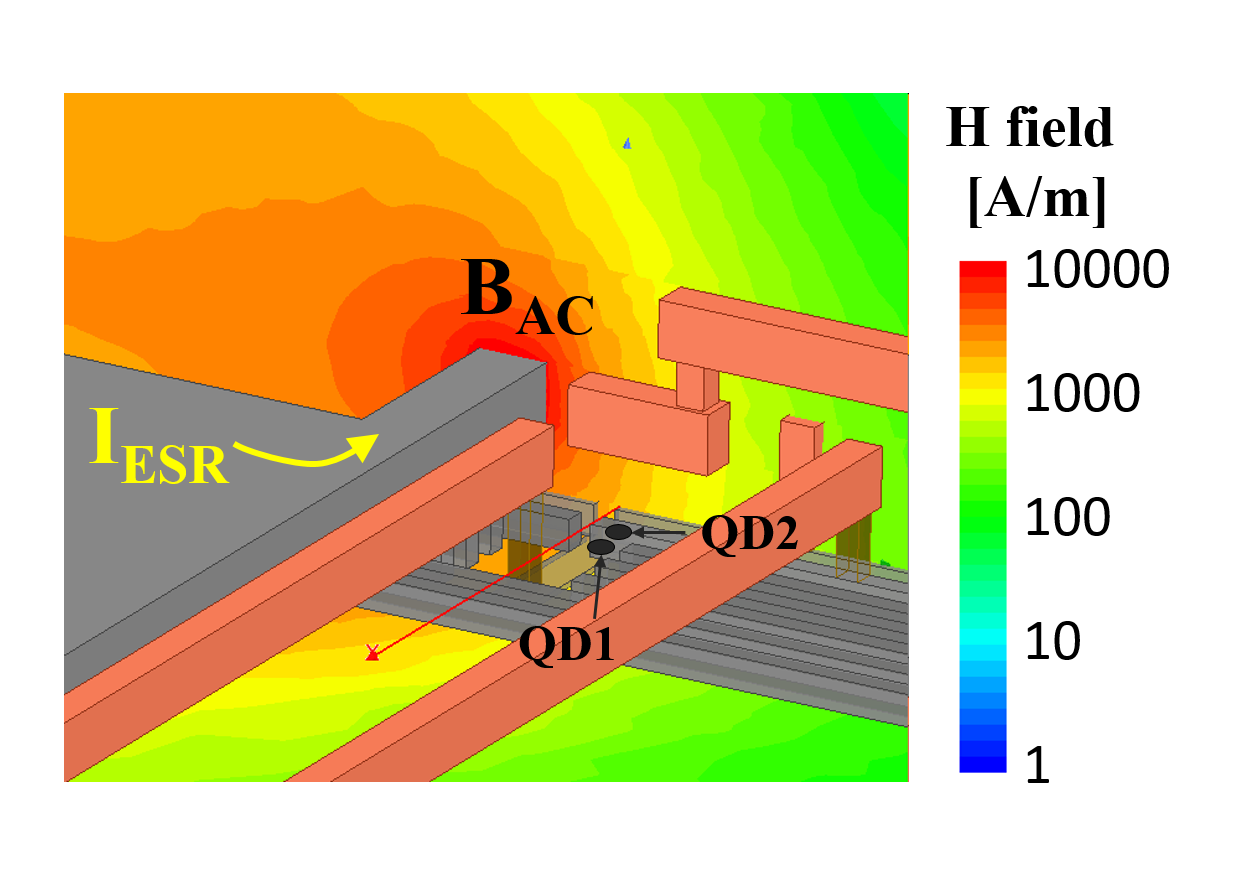}
\caption{Cross-sectional view in the Si QD2 plane of the magnetic field of the ESR line with two quantum dots at $10\,GHz$ with a $-7\,dBm$ input power, illustration from \cite{niquet_challenges_2020}.}
\label{fig_B_ESR_dum}
\end{figure}

The simulated structure is described in Fig.\,\ref{fig_ESR_dum} and \ref{fig_B_ESR_dum}.
\subsection{ESR line configuration impact}
\begin{figure}[!h]
\centering
\includegraphics[width=0.85\columnwidth]{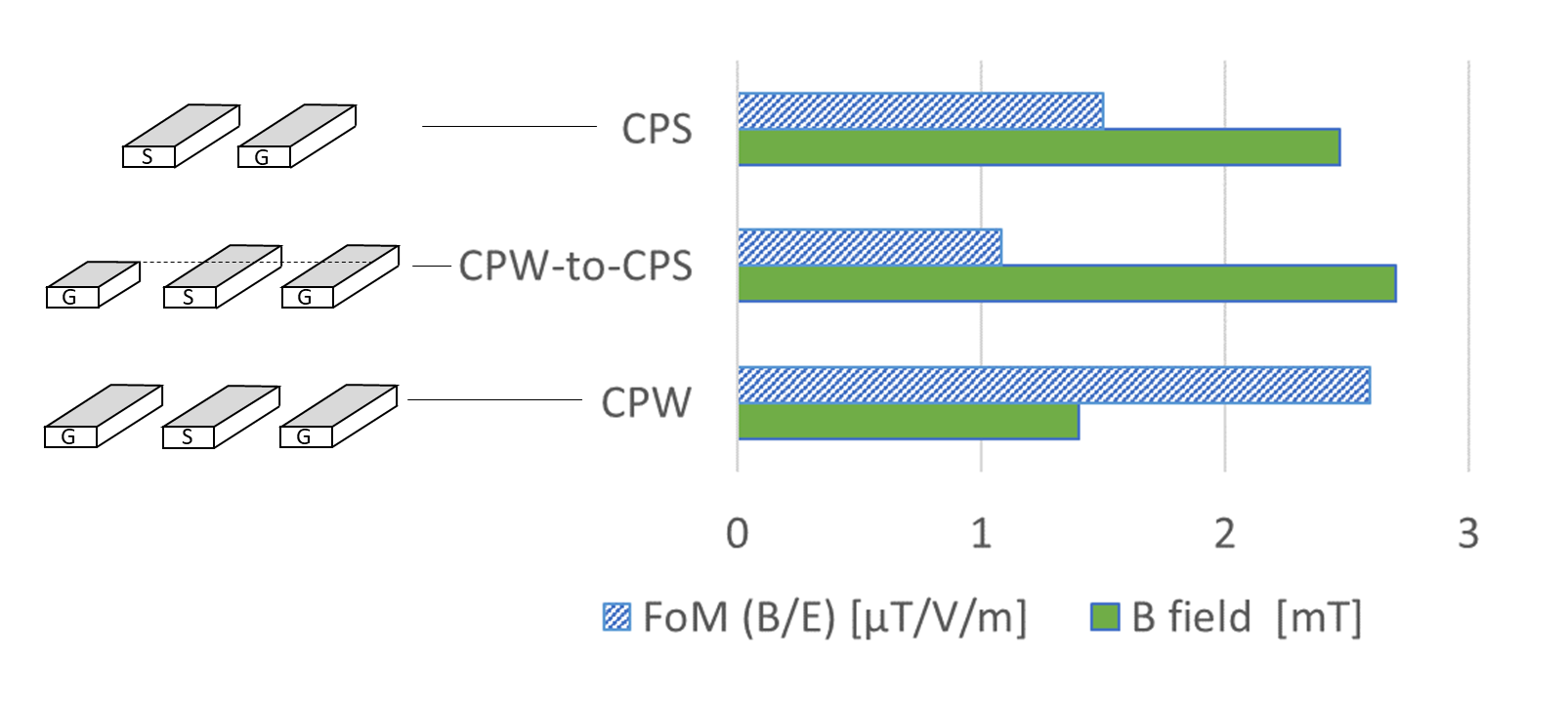}
\caption{ESR line configurations and their impact on FoM, average fields at QDs locations, $Pin$=$-7\,dBm$. The bottom scale is common to all quantities.}
\label{fig_FOM_geo}
\end{figure}

The transmission line configuration of the ESR line is typically a coplanar stripline (CPS) or a coplanar waveguide (CPW) or a coplanar-to-stripline using a balun (CPW-to-CPS) \cite{dehollain_nanoscale_2013}, terminated by a short-circuit placed near the Qubit QDs, where the magnetic field $B$ and the ratio $B$/$E$ are to be maximized. Figure\,\ref{fig_FOM_geo} compares the FoM of the different ESR line configurations and points out the trade-off between maximal $B$ field and maximal $B$/$E$ ratio. The CPW configuration should be chosen if $E$ field is an issue. However, this configuration has the disadvantage of having a $B$ field divided by two as it has two return paths for electrical current, contrary to the two other configurations (Fig.\,\ref{fig_FOM_geo}).
\subsection{Stack impact}
\begin{figure}[!h]
\centering
\includegraphics[width=1\columnwidth]{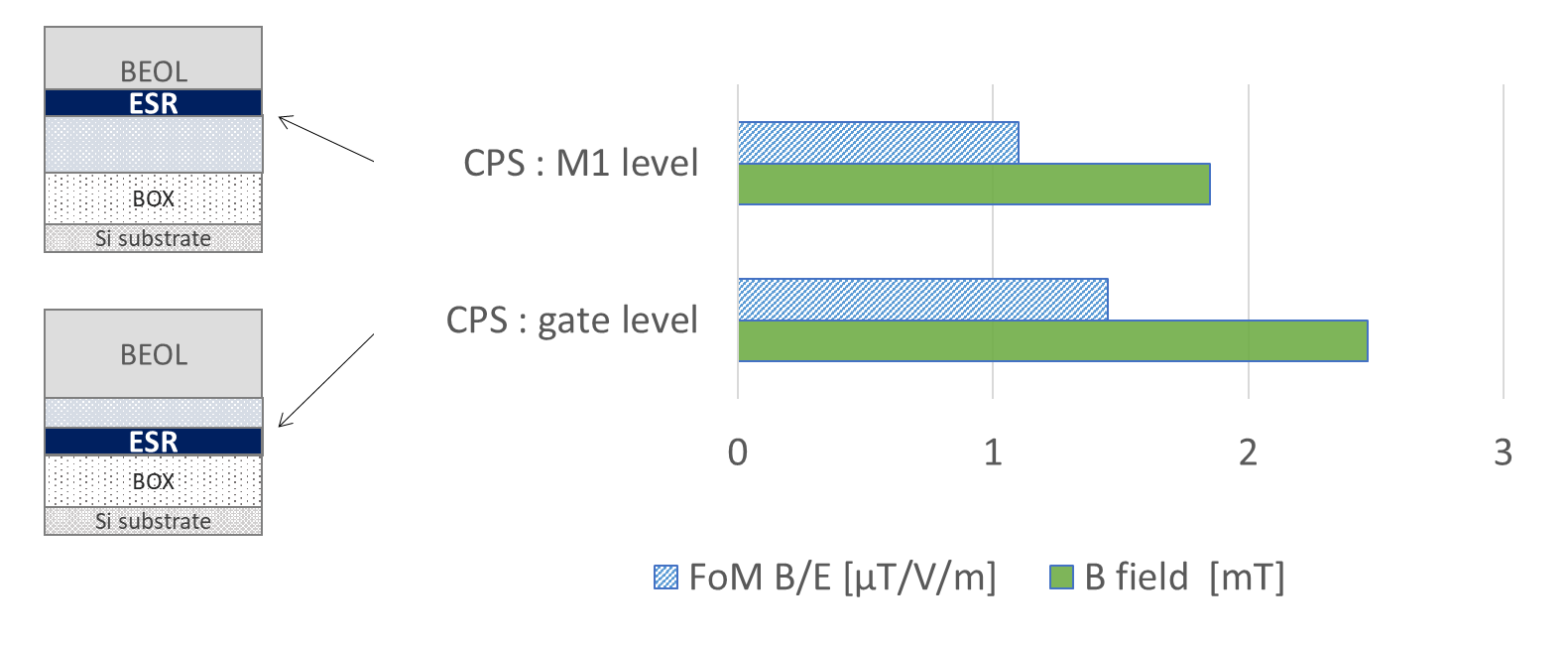}
\caption{BEOL stacks and their impact on ESR line FoM, average fields at QDs locations, $Pin$=$-7\,dBm$. The bottom scale is common to all quantities.}
\label{fig_FOM_stack}
\end{figure}

The impact of the stack for designing the ESR line is studied considering two cases: either the fabrication of the CPS line at the gate level or at the first metallization level M1 (Fig.\,\ref{fig_FOM_stack}).
The interpretation of the results is straightforward: the decrease of $B$/$E$ when using M1 level is due to the decrease of $B$ field as the distance between the QDs and the nano-antenna is larger.

\subsection{EM environment and positioning impact}
Simulation results show that taking into account all the conductive, dielectric layers and polysilicon gates and dummies has a strong effect on the electric field, making it much more inhomogeneous along the line, contrary to the usually simulated simple ESR line geometry evaluation.

Dummy shapes are usually added because a certain metal density is required to comply with foundries density design rule checks. Their main purpose is to improve planarity for manufacturing. In advanced technologies, they can also address issues associated with stress, rapid thermal annealing, and etch.

\begin{figure}[!h]
\centering
\includegraphics[width=0.9\columnwidth]{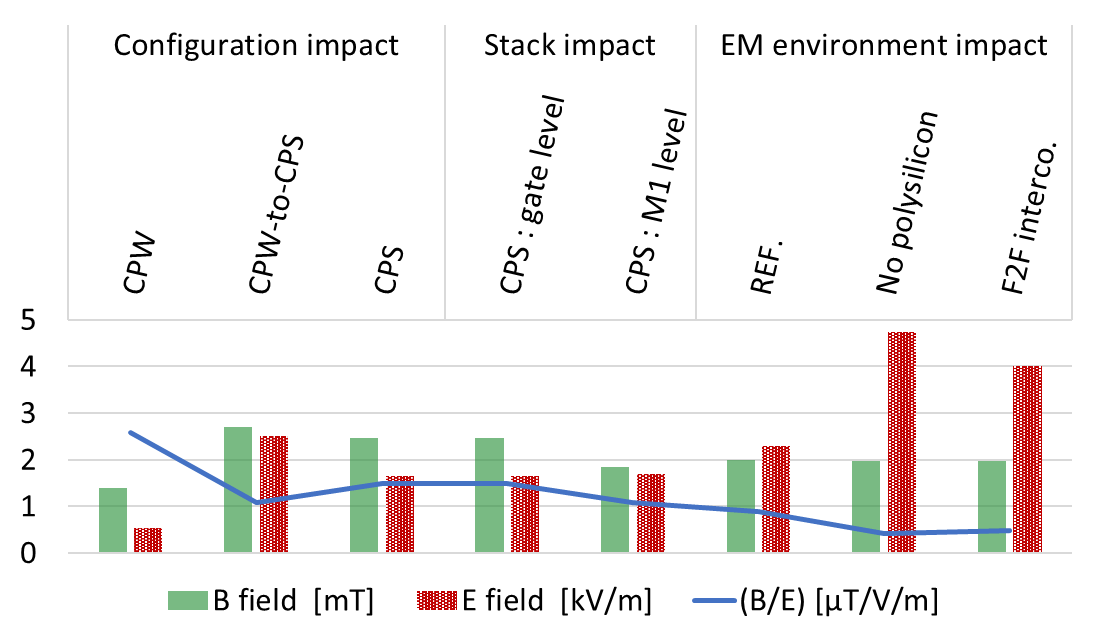}
\caption{Summary of impact on ESR line FoM, average fields at QDs locations, $Pin$$=$$-7\,dBm$ (REF.: CPW-to-CPS at M1 level, with polysilicon gates and dummies and with no F2F interconnect). The left scale is common to all quantities.}
\label{fig_FOM_summary}
\end{figure}

For the ESR line represented in Fig.\,\ref{fig_ESR_dum}, the polysilicon gates and dummies of few nanometers can reduce up to $75\,\%$ the $E$ field thanks to their screening effect.
Inversely, interconnect network of the QDs can increase the electric field locally and degrade the $B/E$. 
When connecting the QDs with exchange gates in a face-to-face configuration \cite{bedecarrats_new_2021}, extra gate interconnect can lead to extra $E$ field.
Thus, a precise multi-scale description of the device and its EM environment in the simulation platform has to be added to the usual ESR stand-alone device evaluation for accurate FoM assessment, as summarized in Fig.\,\ref{fig_FOM_summary}. 
\section{Comparison with experimental data}
A co-design ‘ESR line/qubit’ using a dedicated state-of-the-art CMOS FDSOI technology \cite{bedecarrats_new_2021} has been fabricated and characterized at room and cryogenic temperatures using a Vector-Network-Analyzer with standard and on-chip calibration.
For the 28FDSOI conductive layers, we have used their RRR (Residual-resistance ratio) values based on $4\,K$ experimental results and have adjusted the conductivity for the nano-antenna part (Fig.\,\ref{fig_ESR_dum}) in M1 to $3.10^7\,S/m$.

\begin{figure}[!h]
\centering
\includegraphics[width=0.85\columnwidth]{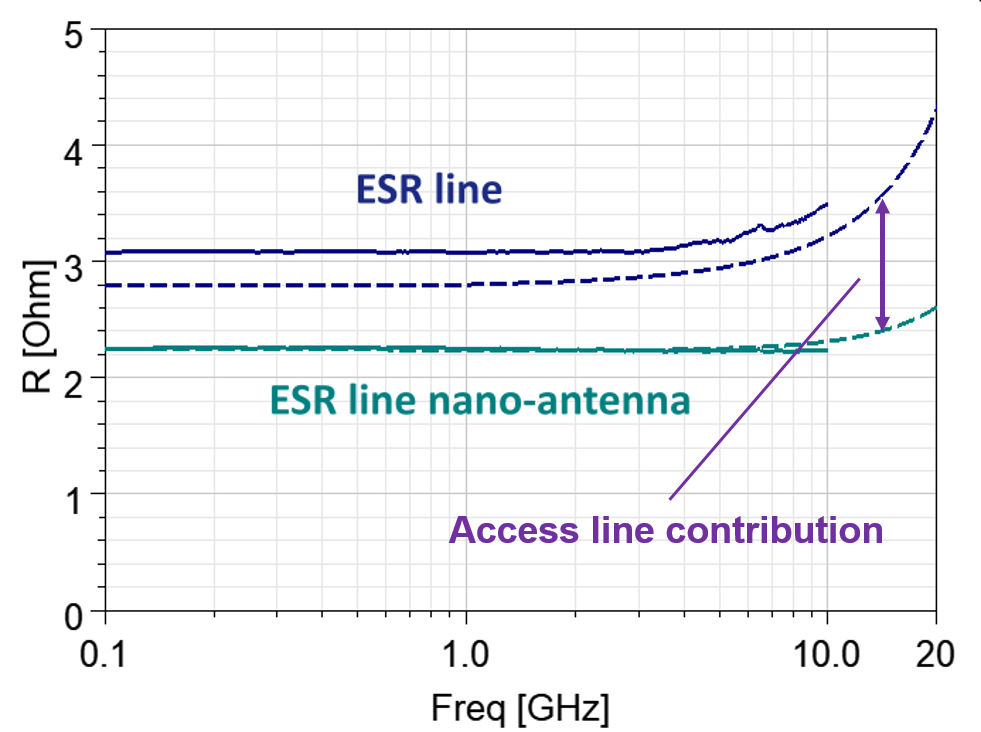}
\caption{Comparison of the ESR line overall resistance (including access lines) with the ESR nano-antenna resistance at cryogenic temperature in the [$100\,MHz$ – $20\,GHz$]: experimental results (lines) and EM simulations (dashed lines).}
\label{fig_R_contrib}
\end{figure}

Fig.\,\ref{fig_R_contrib} shows the simulation and experimental results of both the overall ESR line resistance and the de-embedded nano-antenna part of the ESR line. The nano-antenna resistive part of the ESR line is obtained using some dedicated RF de-embedding test structures, and represents over $60\,\%$ of the total resistance up to $10\,GHz$ demonstrating the low impact of the access line resistance. Therefore, the resistive losses in the ESR line are mainly attributed to the M1 resistivity of the nano-antenna.
In addition, Fig.\,\ref{fig_R_contrib} highlights the quasi-static behaviour of the nano-antenna (quasi-constant resistance over frequency) as a consequence of a very high wavelength to geometric length ratio.

\begin{figure}[!h]
\centering
\includegraphics[width=0.95\columnwidth]{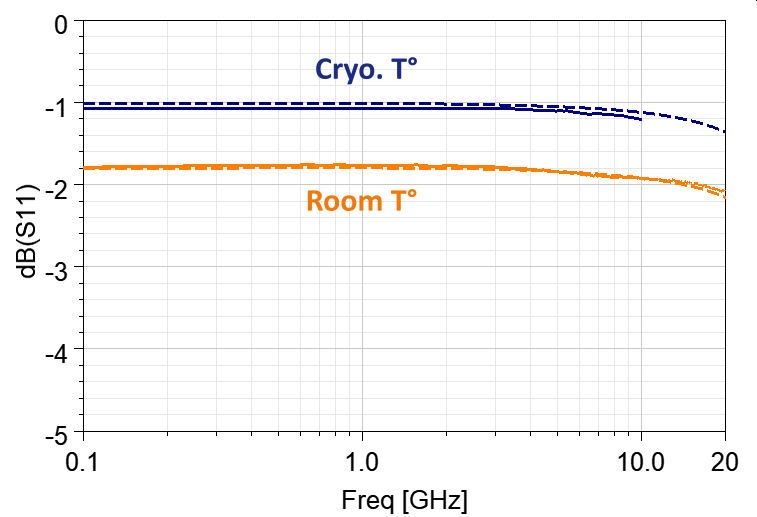}
\caption{Comparison of experimental results (lines) with EM simulations (dashed lines) S parameters in the [$100\,MHz$ – $20\,GHz$] of a co-design ‘ESR line/qubit’ using a double quantum dots with exchange gate in a state-of-the-art CMOS FDSOI technology using a dedicated FEOL and a simplified BEOL in 28FDSOI.}
\label{fig_S11_simu_carac}
\end{figure}

As illustrated in Fig.\,\ref{fig_R_contrib} and \ref{fig_S11_simu_carac}, CEA-LETI simulation platform gives very good agreement between measurements and simulations over a wide frequency range up to $20\,GHz$. 
Moreover, the obtained results in Fig.\,\ref{fig_S11_simu_carac} show both the wideband and low-loss characteristics of the ESR line, with return loss parameter (S11) advantageously reduced at cryogenic temperature.
And the reduction of S11 parameter at cryogenic temperature is mainly due to higher conductivities of the BEOL of the ESR line, in particular that of the nano-antenna.
\section{Conclusion}
Evaluation of the ESR line control EM fields with QDs is performed using a dedicated simulation platform. While only ESR line geometry impact had been studied up to now, we also include in this study the technological stack and the EM environment, considering dummies and interconnects at the vicinity of the QDs, and simulations results clearly indicate their significant impact.
Finally, this simulation platform being experimentally validated, it can be used as a predictive tool to co-design ESR line and QDs and to explore new materials like superconductors for control efficiency optimization.

\section{Acknowledgment}
We acknowledge Spintec and Lateqs laboratories and especially Laurent Vila and Cécile Grezes for their technical support with the experimental setup. 
\section{Fundings}
This work was partly supported by the EU through the H2020 QLSI project and the European Research Council (ERC) Synergy QuCube project.
\bibliography{mybibfile}

\end{document}